\documentclass[preprint]{aastex631}
\usepackage{natbib,graphics}

\begin{document}
\title{Evolution of a Mode of Oscillation Within Turbulent Accretion Disks}
\author{Robert V. Wagoner\altaffiliation{1}  and Celia R. Tandon}
\affil{Dept.\ of Physics and KIPAC, Stanford University}
\altaffiliation{wagoner@stanford.edu}

\begin{abstract}
We investigate the effects of subsonic turbulence on a normal mode of oscillation [a possible origin of the high-frequency quasi-periodic oscillations (HFQPOs) within some black hole accretion disks]. We consider perturbations of a time-dependent background (steady state disk plus turbulence), obtaining an oscillator equation with stochastic damping, (mildly) nonlinear restoring, and stochastic driving forces. The (long-term) mean values of our turbulent functions vanish. In particular, turbulence does not damp the oscillation modes, so `turbulent viscosity' is not operative. However, the frequency components of the turbulent driving force near that of the mode can produce significant changes in the amplitude of the mode. Even with an additional (phenomenological constant) source of damping, this leads to an eventual `blowout' (onset of effects of nonlinearity) if the turbulence is sufficiently strong or the damping constant is sufficiently small. The infrequent large increases in the energy of the mode could be related to the observed low duty cycles of the HFQPOs. The width of the peak in the power spectral density (PSD) is proportional to the amount of nonlinearity. A comparison with observed continuum PSDs indicates the conditions required for visibility of the mode.
\end{abstract}

\section{Introduction}

Consider ideal Newtonian hydrodynamics \citep{tb}. This is a useful first approximation in the following exploratory analysis of the interaction of a normal mode of oscillation with turbulence. Although the initial application will be to (geometrically thin) black hole accretion disks, the effects of general relativity should not change the general nature of our results (due to the Principle of Equivalence, within the small volume of the mode). The effects of the magnetic fields within such disks on the mode are unclear, and will be discussed in the final Section.

The changes that we find in the energy of the mode are produced mainly by the turbulent driving force. This amplification may be relevant to the high-frequency quasi-periodic oscillations (HFQPOs) observed in some of the stellar mass to supermassive black hole (Figure 1) sources \citep{smi,stw}. They typically have stable frequencies, but low duty cycles. The opposite is the case for the black hole low-frequency QPOs and most of the QPOs of accreting neutron stars \citep{rm06}. 

\begin{figure}[tbp]
\plotone{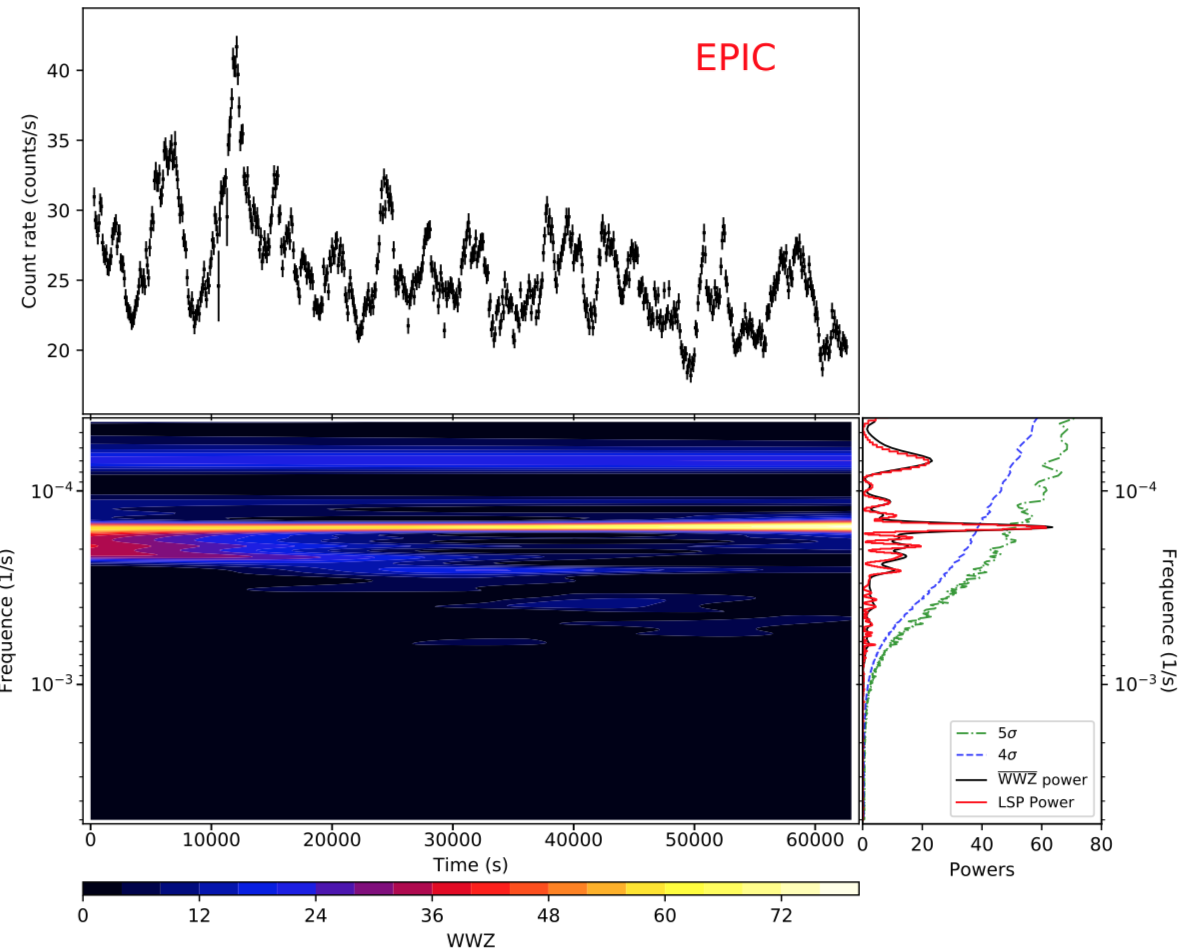}
\caption{An X-ray light curve (top) and power spectral density (right, showing the HFQPO with a period of 1.8 hours) of the narrow line Seyfert 1 active galactic nucleus  Mrk766, powered by a black hole of mass $7\times 10^6$ solar masses \citep{zh}.}  
\end{figure}

Our major focus is the evolution of the amplitude of the mode. Previous investigations of oscillators subject to stochastic damping \citep{go,osad} have mainly considered time averages, and took the damping to have a positive mean value.

We take the background model of the accretion disk to be composed of two components: a) the stationary and axisymmetric thin disk, and b) the contribution of subsonic turbulence [continuously generated by the strong magneto-rotational instability, which transfers the large free energy of the rate of shear ($rd\Omega/dr$) in the disk into the turbulent cascades \citep{bas}]. Thus the background velocity, pressure, and mass density are decomposed as
\begin{equation}
\vec{u}  = r\Omega(r)\vec{e} _\phi + \vec{v} (turb.)  \ , \:
p = p_0(r,z) + p_1(turb.)  \ , \:  \rho = \rho_0(r,z)  + \rho_1(turb.)  \ ,   \label{eq:background}
\end{equation} 
with $|\nabla\cdot\vec{v}| \ll |\nabla\times\vec{v}|$ and the turbulent Mach number $\mathcal{M} \equiv v/c_s  < 1$, where $c_s$ is the speed of sound.  

We consider an adiabatic perturbation of this background from a single normal mode, and neglect mode-mode interactions. We also assume that the turbulence is unaffected by the mode. Then the displacement vector of the mode is taken to be of the form 
\begin{equation}
\vec{\xi} = A(t)\vec{\hat{\xi}} (\vec{x})  \ ,   \label{eq:mode}
\end{equation}
where the eigenfunction $\vec{\hat{\xi}}$ (taken to be axisymmetric, a property of many of the most observable modes) is approximated as that calculated from the zero-order steady background model, and the amplitude $A$ is dimensionless. The density fluctuation produced by the mode is $\Delta\rho/\rho \cong -\nabla\cdot\vec{\xi} \sim A$. The perturbation theory that we shall now employ therefore requires that $A^2 \lesssim 1$.

\section{Evolution of the Mode}

To analyze the dynamics of the mode, we employ the approach of \citet{sch} (especially Appendix I). With $ d/dt = \partial /\partial t + \vec{u}\cdot\nabla $, and allowing for the time dependence of the unperturbed background model, equations (I24) and (I25) of \citet{sch} give the components  
 \begin{equation}
\frac{d^2 \xi_i}{dt^2} = \frac{\partial^2 \xi_i}{\partial t^2}  + 2\vec{u}\cdot\nabla\frac{\partial\xi_i}{\partial t}  + \left(\frac{\partial\vec{u}}{\partial t}\cdot\nabla\right)\xi_i + \vec{u}\cdot\nabla(\vec{u}\cdot\nabla\xi_i)  \label{eq:accel}
 \end{equation} 
 of the acceleration of the perturbation. 
 We invoke the Cowling approximation (neglecting perturbations of the gravitational potential $\Phi$, since the mass of the disk is much less than that of the black hole), and keep the first and second order perturbations of the pressure gradient restoring force. In addition, we add to the equation of motion [I37 of \citet{sch}] the main external driving force per unit volume, the divergence of the turbulent Reynolds stress tensor. 
 
 This then gives the full equation of motion 
 \begin{eqnarray}
\rho d^2\xi_i/dt^2 & = & - (\nabla\cdot\vec{\xi}) \nabla_i p + \nabla_i (\vec{\xi}\cdot\nabla)p + \nabla_i (p\Gamma_1\nabla\cdot\vec{\xi}) - \rho(\vec{\xi}\cdot\nabla)\nabla_i\Phi(r,z) - \nabla^j(p Q_{ij})  \nonumber  \\
                            &    & - (1/2)\rho \xi^j \xi^k \nabla_j \nabla_k \nabla_i \Phi(r,z) - \nabla^j(\rho v_i v_j) \ ,   \label{eq:force}
\end{eqnarray} 
where $\Gamma_1$ is the adiabatic index of the perturbed pressure. The quantity $Q^i_j = k_1 \Theta^i_j + k_2 \Theta\delta^i_j + k_3 \Xi^i_j + k_4 \Xi\delta^i_j$, where $| k_n | \sim 1$, with 
$\Theta^i_j = \nabla_k \xi^k \nabla_j \xi^i $ and $\Xi^i_j = \nabla_k \xi^i \nabla_j \xi^k $ . 

Operating with $\int d^3x \hat{\xi^i}$ on the above equation then gives
\begin{equation}
c_0 d^{2}A/dt^2 + f_1(t) dA/dt + [g_0 + g_1(t)] A + [n_0 + n_1(t)] A^2 = f_2(t) \ ,  \label{eq:projection}
\end{equation} 
with the time-dependent functions generated by the turbulence. (Interactions with other modes would also contribute to the term $g_1(t)$, proportional to their amplitudes.) The constant $c_0 = \int d^3x \rho\hat{\xi}^i \hat{\xi}_i $ and the constant $g_0 = c_0\omega_0^2$, where $\omega_0(\sim \Omega)$ is the eigenfrequency of the mode in the absence of turbulence and the nonlinearity in the restoring force. Employing the phase $\tau\equiv\omega_0 t$ and dividing the above equation by $c_0$ gives our master equation
\begin{equation}
d^2 A/d\tau^2 + F_1(\tau)dA/d\tau + [1 + G_1(\tau) ] A  + [N_0 + N_1(\tau)] A^2 = F_2(\tau) \ ,    \label{eq:master}  
\end{equation} 
in which all quantities are dimensionless. From equation~(\ref{eq:force}), 
it is seen that $| N_0 | \sim | \nabla_i \hat{\xi}^j | \sim 1$.

Multiplying equation~(\ref{eq:master}) by $dA/d\tau$ gives
\begin{equation}
\frac{dE}{d\tau} = F_2(\tau)\frac{dA}{d\tau} - F_1(\tau)\left(\frac{dA}{d\tau}\right)^2 
 - \frac{1}{2} G_1(\tau)\frac{dA^2}{d\tau} - \frac{1}{3} N_1(\tau) \frac{dA^3}{d\tau}  \ . \label{eq:evolution}
\end{equation}
We have introduced the (dimensionless) energy 
\begin{equation}
E = \frac{1}{2}\left(\frac{dA}{d\tau}\right)^2 +  \frac{1}{2}A^2 + \frac{1}{3} N_0 A^3   \label{eq:energy}
\end{equation}
of the mode. 

We find that  
\begin{equation}
F_1  =  \frac{2}{c_0\omega_0}\int d^3x\rho\hat{\xi}^i (\vec{u}\cdot\nabla)\hat{\xi}_i =  \frac{1}{c_0}\frac{d}{d\tau}\int d^3x \hat{\xi}^2 \rho_1 \equiv dH/d\tau \ . \label{eq:F1temp}
\end{equation}
To obtain the final expression, we have employed an integration by parts and the conservation of mass [$\partial \rho/\partial t + \nabla\cdot(\rho\vec{u}) = 0$]. We see that $F_1$ is proportional to the rate of change of the mass within the mode, so a decreasing mass produces growth of the mode. However, its long time average $\langle F_1\rangle = 0$. Therefore, we now include a phenomenological source of damping, so that
\begin{equation}
F_1 = dH/d\tau + D_* \: ,   \label{eq:F1}
\end{equation}
with the constant $D_* > 0$. 

The function $g_1(t)$ is generated by the last two terms in equation~(\ref{eq:accel}) and the first three terms on the right-hand-side of equation~(\ref{eq:force}) (which are larger than the fourth). The function $n_1(t)$ is  generated by the term in equation~(\ref{eq:force}) involving $Q^i_j$. Therefore, 
\begin{equation}
|G_1|  \sim |N_1| \sim  |p_1|/p_0  \ .    \label{eq:G1N1}
\end{equation}
Finally, we obtain 
\begin{equation}
F_2  =  \frac{1}{c_0\omega_0^2}\int d^3x\hat{\xi}^i\mathcal{F}_i = \frac{1}{c_0\omega_0^2}\int d^3x\rho_{0}v^i (\vec{v}\cdot\nabla)\hat{\xi}_i \,  ,   \label{eq:F2} 
\end{equation} 
where $\mathcal{F}_i = -\nabla^j(\rho_0 v_i v_j)$ is the turbulent driving force per unit volume. 

Consider briefly the case when $A^2 \ll 1$, so that we can neglect the nonlinear term in equation~(\ref{eq:master}). Employing the the change of variable $A(\tau) = a(\tau) \exp[-0.5\int_{\tau_0}^{\tau} F_1(s)ds]$ and the upper limits on the magnitudes of the stochastic functions (discussed in the next section)  when $\mathcal{M} \ll 1 $ (giving $| G_1 + (1/2)dF_{1}/d\tau + (1/4)F_{1}^{2} | \ll 1$), we obtain $d^{2}a/d\tau^2 + a = F_2(\tau)\exp[0.5\int_{\tau_0}^{\tau} F_1(s)ds]$ from equation~(\ref{eq:master}). Employing the relevant Green's function, we then obtain the solution 
\begin{eqnarray}
A(\tau) &  =  &  \cos\tau\{A_0\exp[-\onehalf\int_0^{\tau} F_1(s)ds] - 
\int_0^\tau \sin\tau' F_2(\tau')\exp[-\onehalf\int_{\tau'}^{\tau} F_1(s)ds] d\tau' \} \nonumber \\
           &   +  &  \sin\tau\{\dot{A}_0\exp[-\onehalf\int_0^{\tau} F_1(s)ds] + 
\int_0^\tau \cos\tau' F_2(\tau')\exp[-\onehalf\int_{\tau'}^{\tau} F_1(s)ds]d\tau' \} , \label{eq:linear}
\end{eqnarray}
corresponding to the initial conditions $A = A_0$ and $dA/d\tau = \dot{A}_0$, at $\tau = \tau_0 = 0$.

Let us consider the evolution of $A(\tau)$ at late times ($\langle F_1 \rangle\tau\gg 1$), with $\langle F_1 \rangle\equiv (\tau - \tau')^{-1}\int_{\tau'}^{\tau} F_1(s)ds \cong D_* \ll 1$ . From equation~(\ref{eq:linear}) we then obtain (until effects of nonlinearity become important)
\begin{equation}
  A(\tau) \approx \int_{\tau - \tau_*}^{\tau} F_{2}(\tau')\sin(\tau - \tau') d\tau' = \int_{0}^{\tau_*} F_{2}(\tau - x)\sin(x) dx  \, , \label{eq:late}
\end{equation}
where $\tau_* \equiv 1/D_*$ and $\tau \gg \tau_* \gg 1$. 

Including the mild nonlinearity, the evolution of the energy $E(\tau)$ of the mode is governed by equation~(\ref{eq:evolution}). From equation~(\ref{eq:energy}), the potential energy is $PE = A^{2}/2 + N_{0}A^{3} /3$. $|A|$ becomes unbounded if the energy reaches $1/(6N_0^2)$, which occurs when $A = -1/N_0$, as shown in Figure 2. Will equilibrium be achieved before this blowout can occur? (Of course, `blowout' only indicates that nonlinearities have become important, so the evolution cannot be accurately continued. The energy would remain bounded if the coefficient of an $A^4$ term was positive.)  Our observational predictions (in Section 4) are all based upon choices of parameters that do not produce a blowout.

\begin{figure}[tbp]
\plotone{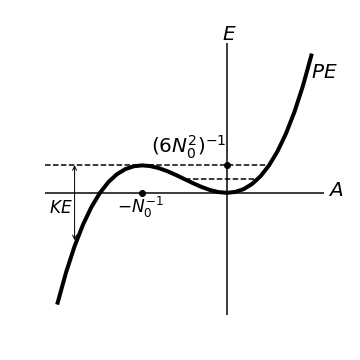}
\caption{The potential energy $PE = (1/2)A^2 + (1/3)N_0 A^3$, illustrating the condition for a `blowout' of the kinetic energy (KE) (onset of the effects of nonlinearity).}  
\end{figure}

Averaging over a time interval $\delta\tau$ (with $2\pi \ll \delta\tau \ll \tau$), equation~(\ref{eq:evolution}) gives 
\begin{equation}
\langle dE/d\tau \rangle \cong \langle F_2 (dA/d\tau) \rangle - D_* \langle (dA/d\tau)^2 \rangle \: . \label{eq:average}
\end{equation}
(We have found that the contributions of $G_1$ and $N_1$ are negligible.)
At blowout, $\langle (dA/d\tau)^2 \rangle \cong \langle A^2 \rangle \cong 1/(2N_0^2)$. Initially, the driving term will be larger than the damping term, since it is only proportional to $A$. Therefore, the condition for equilibrium to be achieved before a blowout could occur is
\begin{equation}
\langle F_2 dA/d\tau \rangle \lesssim D_* /(2N_0^2)  \; .   \label{eq:blowout}
\end{equation}

\section{Modeling the Turbulence} 

We next consider how to characterize the turbulence. The ensemble and time averages of our four stochastic functions ($H , F_2 , G_1 , N_1$) should vanish. The major properties of a turbulent eddy of radius $\ell$ are its velocity $v_\ell$ and turnover time $t_\ell\sim 2\pi\ell/v_\ell$ . The smallest dimension (radius) of the largest turbulent eddy will be of order the (half) thickness $h(r)$ of the accretion disk. Vertical force balance gives $h\Omega \sim c_s$ . 

We shall consider a normal mode with a similar thickness [such as the fundamental g (also called r) - mode], with period $t_m\sim 2\pi/\Omega \sim t_L$(the period of the largest eddy). \citet{gk} found that an acoustic mode in the Sun couples most strongly to the eddies of the convective turbulence with $t_\ell \lesssim t_m$, giving $\ell \lesssim \mathcal{M}h$ , where $\mathcal{M}$ is now the Mach number of the largest eddies (radius $L\sim h$). We shall assume that the same holds true for the modes in our more strongly turbulent accretion disks \citep{nw}. 
 
The pressure and density fluctuations produced by such eddies are
\begin{equation}
p_1(\vec{x},t) \sim \rho_0 v^2 \ , \:  \rho_1(\vec{x},t) \sim \rho_0 (v/c_s)^2 \ . \label{eq:fluctuations}
\end{equation}
We assume that these fluctuations arise mainly from the conversion of the eddies into acoustic modes \citep{gk}, 
and take $|\nabla|\lesssim 1/h$. A limit on the turbulent velocity comes from the effects of a `turbulent viscosity' on the structure and evolution of accretion disks \citep{kfm}. \citet{bas} find that $\mathcal{M} \approx 0.06 - 0.10$.

We generate the (now dimensionless) eddy lifetimes $\Delta\tau$ (assumed equal to the turnover time $\tau_\ell$) by random sampling a distribution function with mean $\langle\Delta\tau\rangle \lesssim \pi$ , 
and we shall assume no delay between eddies. (Note that the unperturbed period of the mode is $\tau_m = 2\pi$.) We choose a Rayleigh distribution $f(\Delta\tau) = (\Delta\tau/\sigma_*^2) \exp[-(\Delta\tau)^2 /(2\sigma_*^2)] $ . Its mean $\langle\Delta\tau\rangle = \sigma_*\sqrt{\pi/2}$ and its variance is $(2-\pi/2) \sigma_*^2 $. 

From the above equations, it is seen that 
\begin{equation}   
    |H| \sim |G_1| \sim |N_1| \sim |F_2| \sim \eta\mathcal{M}^{2} \; ,    \label{eq:estimates} 
\end{equation}  
where $\eta$ is the ratio of the volume of the dominant correlated eddies within the mode to the volume of the mode. Thus since $\eta\mathcal{M}^2 \ll 1$, we neglect $G_1$ and $N_1$ (compared to unity) in our master equation~(\ref{eq:master}). Also, if $F_1$ is given by equation~(\ref{eq:F1}), we again see that $H(\tau)$ does not affect the long-term solution [equation~(\ref{eq:linear})] to our linear equation, since $\int_{\tau_1}^{\tau_2} F_1(s)ds = [H(\tau_2) - H(\tau_1)] + D_*(\tau_2 - \tau_1) \cong D_*(\tau_2 - \tau_1)$ . 

We approximate the functions $H(\tau)$ and $F_{2}(\tau)$ as
 \begin{equation}
   \Psi_{k}(\tau) = 16Y_{k,n}x^2(1-x)^2 \; , \; x\equiv (\tau - \tau_{n})/\Delta\tau_{n}  \; \;  ( k=1-2, \: 0\leq x\leq 1 ) \label{eq:Psi}
\end{equation}
during each eddy lifetime $\Delta\tau_n$. Note that $|Y_{k,n}|$ is the maximum value of $|\Psi_{k}|$ in eddy $n$ , and the value and first derivative of $\Psi_{k}$ vanish at the beginning and end of the eddy.  

The values of $Y_{k,n}$ are generated by random sampling a Gaussian-Markov conditional probability function $P(Y_{k,n}, \bar{\tau}_n | Y_{k,n-1})$, where $\bar{\tau}_n = (\Delta\tau_{n}+\Delta\tau_{n-1})/2$. Its mean $\langle Y_{k,n}\rangle = \bar{Y} + \exp^{-\bar{\tau}_n/\tau_r}(Y_{k,n-1} - \bar{Y})$, and its variance $\sigma^2_{k,n} = (1-\exp^{-2\bar{\tau}_n/\tau_r}) \sigma_{k}^2$. We choose the equilibrium mean $\bar{Y} = 0$, consistent with our assumption that $\langle\Psi_k\rangle = 0$ for averaging times $\delta\tau \gg \Delta\tau$. 
We choose the relaxation time $\tau_r = K\langle\Delta\tau\rangle$, with the value of $K$ allowing for correlations between subsequent eddies.  We usually choose values of $\sigma_{k} = \eta\mathcal{M}^2 \lesssim 10^{-3}$, consistent with the expectation that $\eta \ll1$ and $\mathcal{M}^2 \lesssim 10^{-2}$.

\section{Results}
\subsection{Evolutions}

For a calculation characterized by particular values of the physical parameters ($\eta\mathcal{M}^2$, $D_*$, $\langle\Delta\tau\rangle$, and $K$), we employ a second-order integrator, Huen's method. We generate M samples of $\Delta\tau$ and $\Psi_{k}(\tau)$. Each evolution has a duration $\tau_{max} \approx M\langle\Delta\tau\rangle$ , with different random realizations of the distributions of $\Delta\tau$ and $Y_k$ during each eddy $n$. The initial conditions are chosen to be $A_0 =\dot{A}_0 = 0$ at $\tau=0$, but the long-term evolution does not depend on them. 

\begin{figure}[tbp]
\plotone{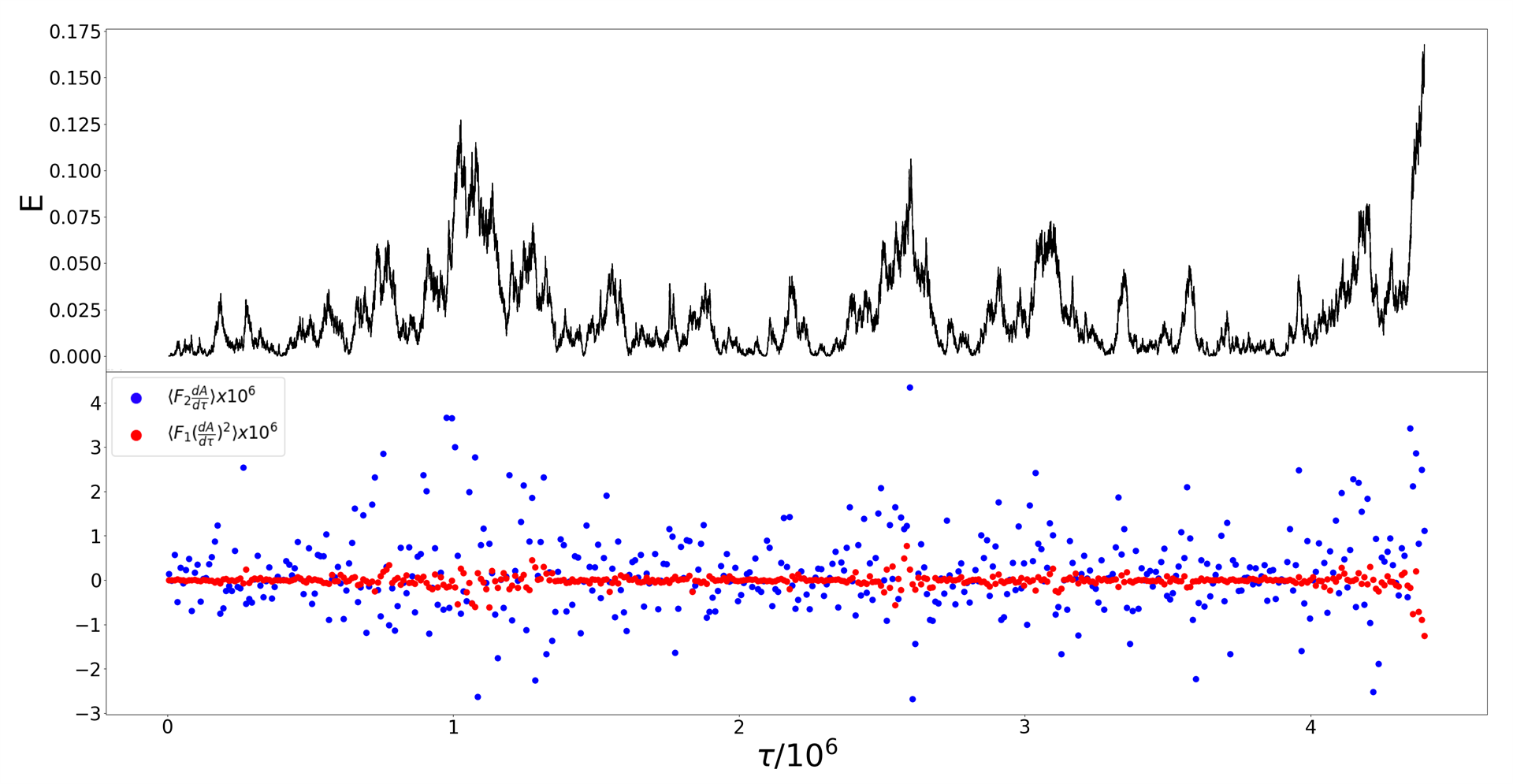}
\caption{An evolution of the energy $E$ of the mode, and that of short term ($\delta\tau = 10^4$) averages of the two stochastic functions [ $F_2 (dA/d\tau)$ and $F_1 (dA/d\tau)^2$ ] which mainly determine it, for $\eta\mathcal{M}^2 = 10^{-3} $, $D_*=  10^{-6}$, $\langle\Delta\tau\rangle = \pi$, and $K = 1$. A blowout occurs at $\tau = 4.4\times 10^6$ .}
\end{figure}

In Figure 3, we see how short term changes in $\langle F_2 dA/d\tau\rangle_{\delta\tau}$ correlate with changes in the energy $E$. The damping function $\langle F_1 (dA/d\tau)^2 \rangle_{\delta\tau}$ is usually subdominant on this averaging time scale, but notice that anti-damping becomes strong just before the blowout. As predicted in Section 2, the energy reaches $E = 1/6$ at blowout for our choice $N_0 = 1$ .

\begin{figure}[tbp] 
\plotone{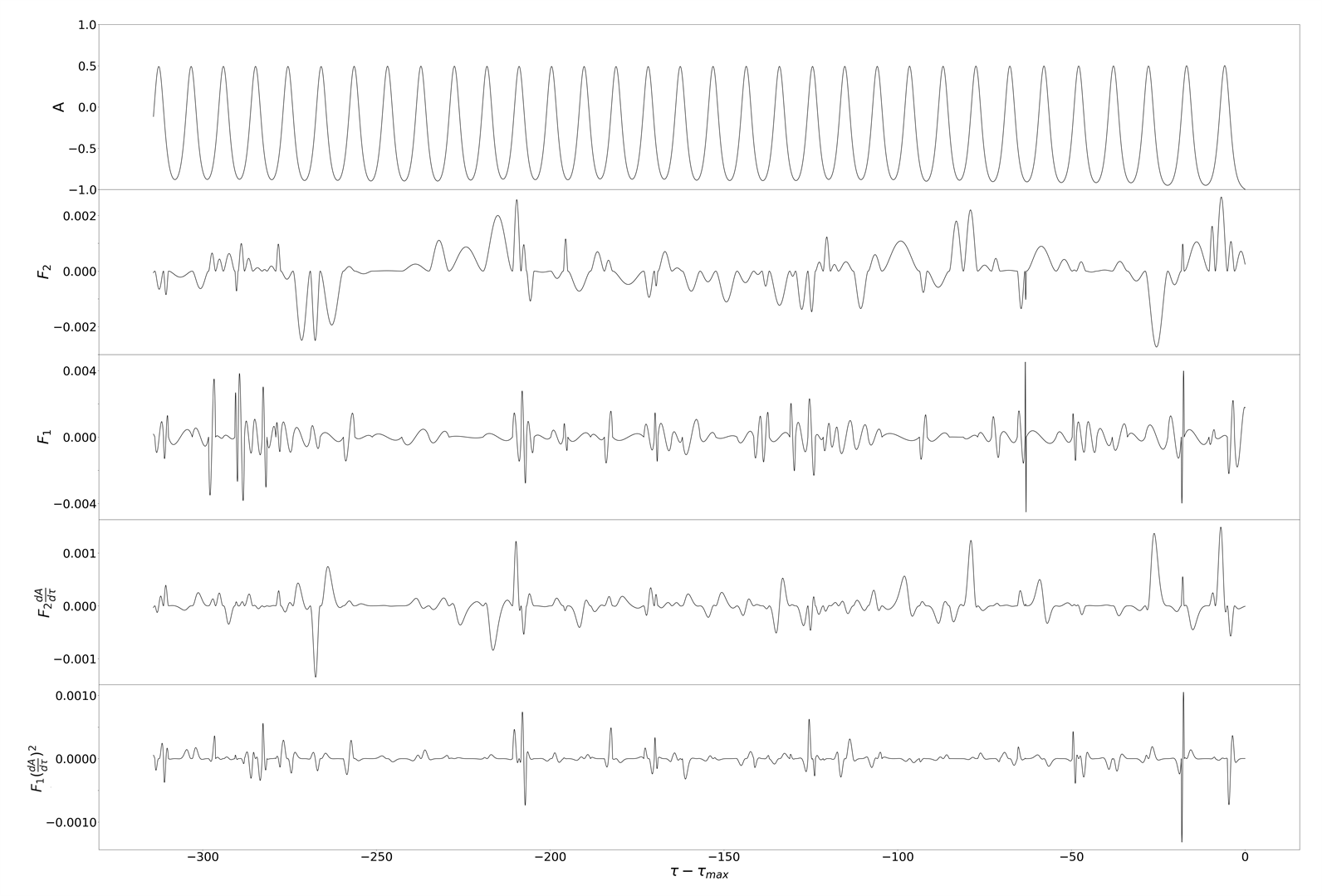}
\caption{A short ($M=10^2$) segment at the end of the evolution of $A$, from the same run as shown in Figure 3, now including $F_2$ and $F_1$. }
\end{figure}

In Figure 4, we see how the negative extent of the amplitude $A$ grows as the blowout is approached, and reaches $A = -1/N_0 = -1$ at blowout. It is also seen that the period of the mode has increased by a factor of about 1.7 since the beginning of the evolution, as also expected from Section 2. Also note the behavior of the driving and damping functions during the eddies. 

\begin{figure}[tbp]
\plotone{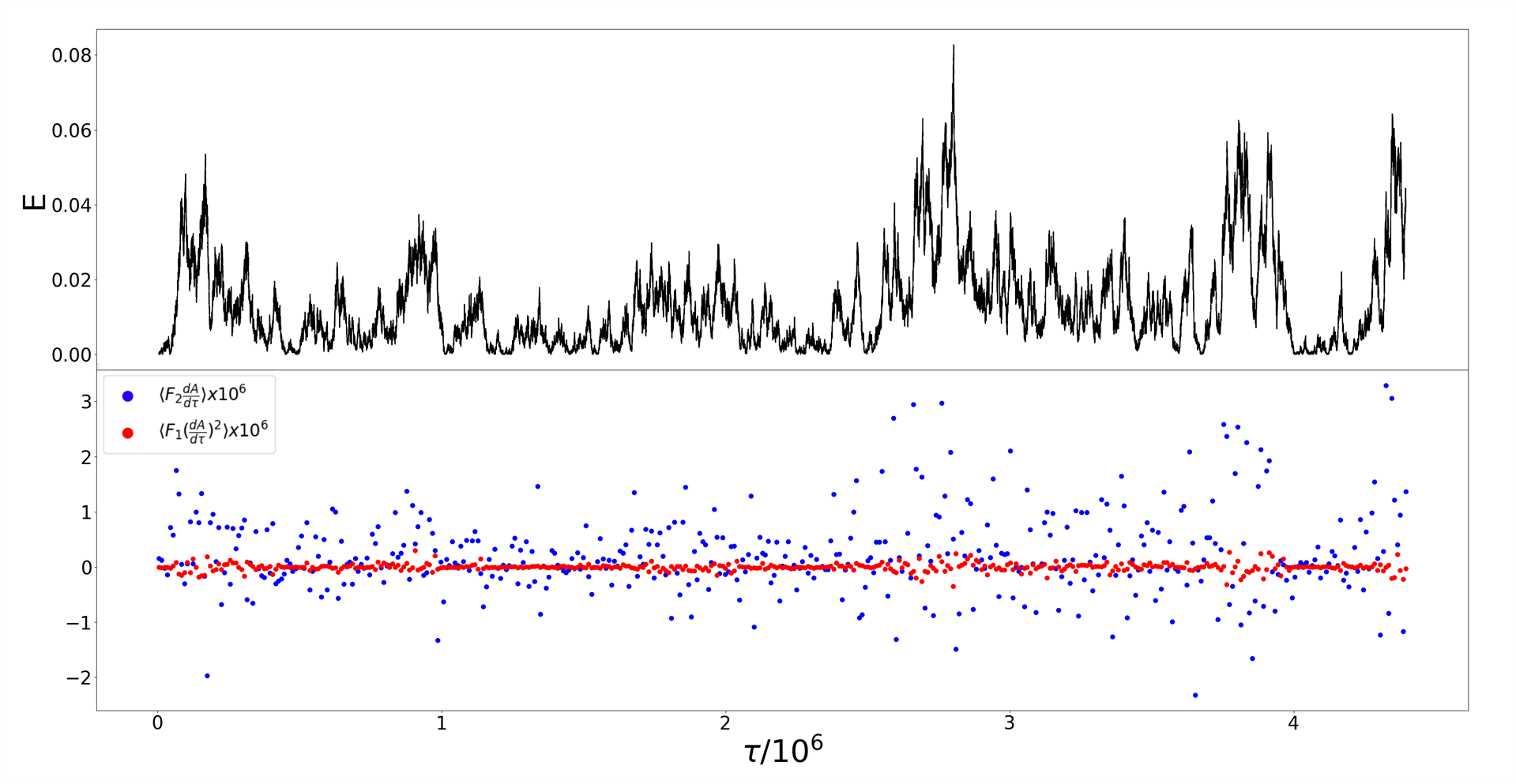}
\caption{Same as Figure 3, but for $D_* = 10^{-5}$. No blowout occurs, at least before the end of this run at $\tau = 2\times 10^7$.}
\end{figure}

In Figure 5, the damping constant has been increased by a factor of 10, which is sufficient to prevent a blowout (at least over the extended time interval indicated). This is our fiducial evolution. The maximum value of the energy (about 1/2 that required for a blowout) is about 10 times larger than its average ( $\langle E \rangle = 8\times 10^{-3}$ ).

\begin{figure}[tbp]
\plotone{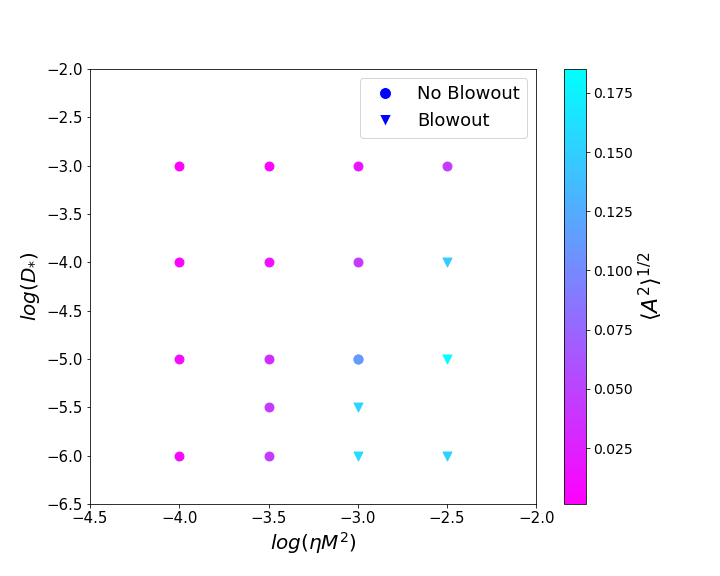}
\caption{The dependence of the r.m.s.\ amplitude of the mode on the driving and damping parameters.}  
\end{figure}

In Figure 6, the parameter choices that produce a blowout or not are indicated, with the corresponding r.m.s.\ 
amplitudes. Consider times $\tau \gg \tau_* = 1/D_* $, so we can employ equation~(\ref{eq:late}) while the effects of nonlinearity are small. Then we obtain from equation~(\ref{eq:average}), when averaging over times $\delta\tau = 10^4 $, 
\begin{equation}
\langle E \rangle (\tau) \cong \int_0^\tau \langle S \rangle (\tau')\exp[D_*(\tau' - \tau)]d\tau' \approx \int_{\tau - \tau_*} ^{\tau}\langle S \rangle (\tau') d\tau' \; , \label{eq:Eevol}  
\end{equation}
where $S(\tau) = F_2(\tau)\int_0^{\tau_*} [dF_2(\tau - x)/d\tau]\sin(x) dx$ . Since $S \propto (\eta\mathcal{M}^{2})^2 $ and the integral spans $\tau_* = 1/D_* $, $\langle E \rangle (\tau) \propto (\eta\mathcal{M}^{2})^2 /D_* $ . Since the blowout occurs at a fixed value of $E=1/6$, the boundary of the blowout region does appear to approximately agree with the dependence $D_* \sim 10(\eta\mathcal{M}^{2})^2$. 

\begin{figure}[tbp] 
\plotone{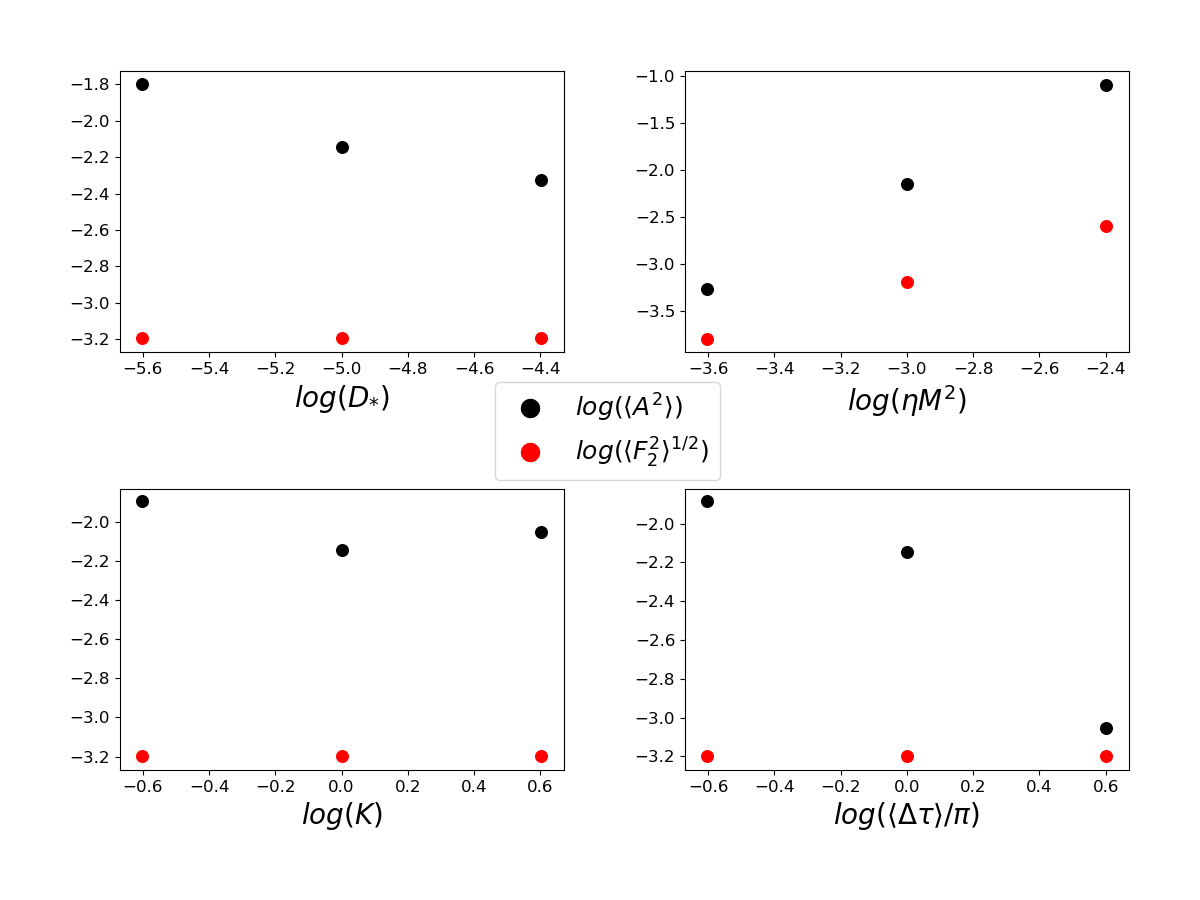}
\caption{The dependence of the values of $\langle A^2 \rangle$ and $\langle F_2^2 \rangle^{1/2}$ on the parameters $D_* $, $\eta\mathcal{M}^2 $, $K$, and $\langle\Delta\tau\rangle $. Averages are over the entire evolution. In each case, the other parameters are at their fiducial value, indicated by their central values on the bottom axes.} 
\end{figure}

From Figure 7, we see how $\langle A^2 \rangle $ and $\langle F_2^2 \rangle^{1/2}$ depend on our four parameters. The dependence $\langle A^2 \rangle \approxeq \langle E \rangle  \sim 0.1D_*^{-1}(\eta\mathcal{M}^2)^2$ is also obtained from this Figure. 

How does the average energy in the mode compare to the average energy in the effective turbulent eddies? The energy in the mode is approximately $\int d^{3}x \rho (\partial\xi^{i}/\partial t) (\partial\xi_{i}/\partial t) \approx c_0\omega^2 A^2 $. The energy in a dominant eddy is approximately $0.5\int d^{3}x \rho v^2  \approx c_0\omega^2 F_2$. Thus equipartition would imply that $\langle A^2\rangle \approx \langle F_2^2\rangle^{1/2}$. The results shown in Figure 7 indicate that this is approximately true for small values of $\eta\mathcal{M}^2$.                                                                  

\subsection{Contribution of the Mode to an Observed Power Spectral Density (PSD)} 

Consider an axisymmetric (the most observable) mode of oscillation at radius $r_m$ with volume $v = 2\pi r_m\zeta h^{2}(r_m)$. Its observed normalized PSD $P_{x}(f)$ ($f=\omega /2\pi$) is related to the the photon count rate $x$ by 
\begin{equation}
  \int P_x(f) df = \langle(\Delta x)^2\rangle / \langle  x\rangle^2 \cong P_{x}(max) \Delta f \: , \label{eq:normPSD}
\end{equation}  
where $\Delta f$ is its full width at half maximum. 
Adopting the approach of \citet{nw}, 
\begin{equation}
x - \langle x\rangle = \Delta x = (\mathcal{F}v/h)(\Delta\rho / \rho) \label{eq:Counts}
\end{equation}
at $r=r_m$. The average photon number flux from the disk is 
$\mathcal{F}(r) = \int_{\nu_1}^{\nu_2} \mathcal{F}_{\nu}(\nu,r) \xi(\nu) d\nu$ 
for a detector efficiency $\xi(\nu)$ . 
We shall employ the approximation  $\mathcal{F}(r) = \mathcal{F}(r_m)(r/r_m)^{-C_1}$. 

In addition, recall that the density fluctuation $\Delta\rho / \rho \approx A(\tau)$, related to its (not normalized) PSD by
\begin{equation}
\int P_A(q)dq = 2\pi\langle A^2\rangle \: .   \label{eq:Pdq}
\end{equation}
The average count rate is (assuming that $C_1 > 2$) 
\begin{equation}
\langle x\rangle = 2\pi\int_{r_{min}}^{r_{max}}\mathcal{F}(r) r dr \approx  2\pi\mathcal{F}(r_m)(C_1 - 2)^{-1} r_m^2 (r_{min}/r_m)^{(2-C_1)} \: ,   \label{eq:xaverage}
\end{equation}
giving 
\begin{equation}
\int P_x(f) df \approx (C_1 - 2)^2 \zeta^2 [h(r_m)/r_m]^2 (r_{min}/r_m)^{2(C_1 - 2)} \langle A^2\rangle \equiv \chi\langle A^2\rangle \: .  \label{eq:fPSD}
\end{equation}

Now consider the observed continuum PSD $P_{x}(\mbox{cont.})$ of black hole sources near the frequency $f_0$ of a HFQPO, with $(M/10 M_\sun)f_0 \sim 200$ Hz for black hole binaries (BHBs) and $\sim 50$ Hz for NLS1 AGNs \citep{stw}. 
We find that $f_0 P_{x}(\mbox{cont.}) \sim (2-20)\times 10^{-4}$ for BHBs and $\sim (3-30)\times 10^{-3}$ for NLS1s.
The corresponding quantity for our mode is given [from equation~(\ref{eq:fPSD})] by
\begin{equation}
f_0 P_{x}(\mbox{max., QPO}) \approx \chi Q \langle A^2 \rangle \; .                          \label{eq:fP}
\end{equation}
The quality factor $Q \equiv f_0/\Delta f$, and  $\chi \sim (\zeta h/r)^2$ . 

During a quasi-equilibrium, when $E(\tau)$ is changing slowly but is not far below its blowout value of $1/6$ (for $N_0^2 =
1$), the effects of our lowest-order nonlinearity control the width of the PSD peak, with the damping and driving forces subdominant. Then the standard analysis \citep{LL} predicts an amplitude-dependent shift of the anharmonic oscillator frequency of 
\begin{equation}
\delta\omega / \omega_0 = \delta q \approx -(5/12)N_0^2 E \; . \label{eq:shift}
\end{equation}
If the energy varies over a range $\Delta E$ during an evolution, it produces a corresponding width $\Delta q$ of the PSD peak. Comparing the range of energies (that occupy a significant fraction of of the time) seen in Figure 5 with the width of the PSD peak corresponding to the same (fiducial) parameters in Figure 9, we see that this relation is approximately valid. It is only indicative, since the effects of higher-order nonlinearities have been neglected. From the results shown in Figure 9, we obtain the approximate relation $\Delta q \sim 1.0  \langle A^2 \rangle^{0.9}$ . 

\begin{figure}[tbp]
\plotone{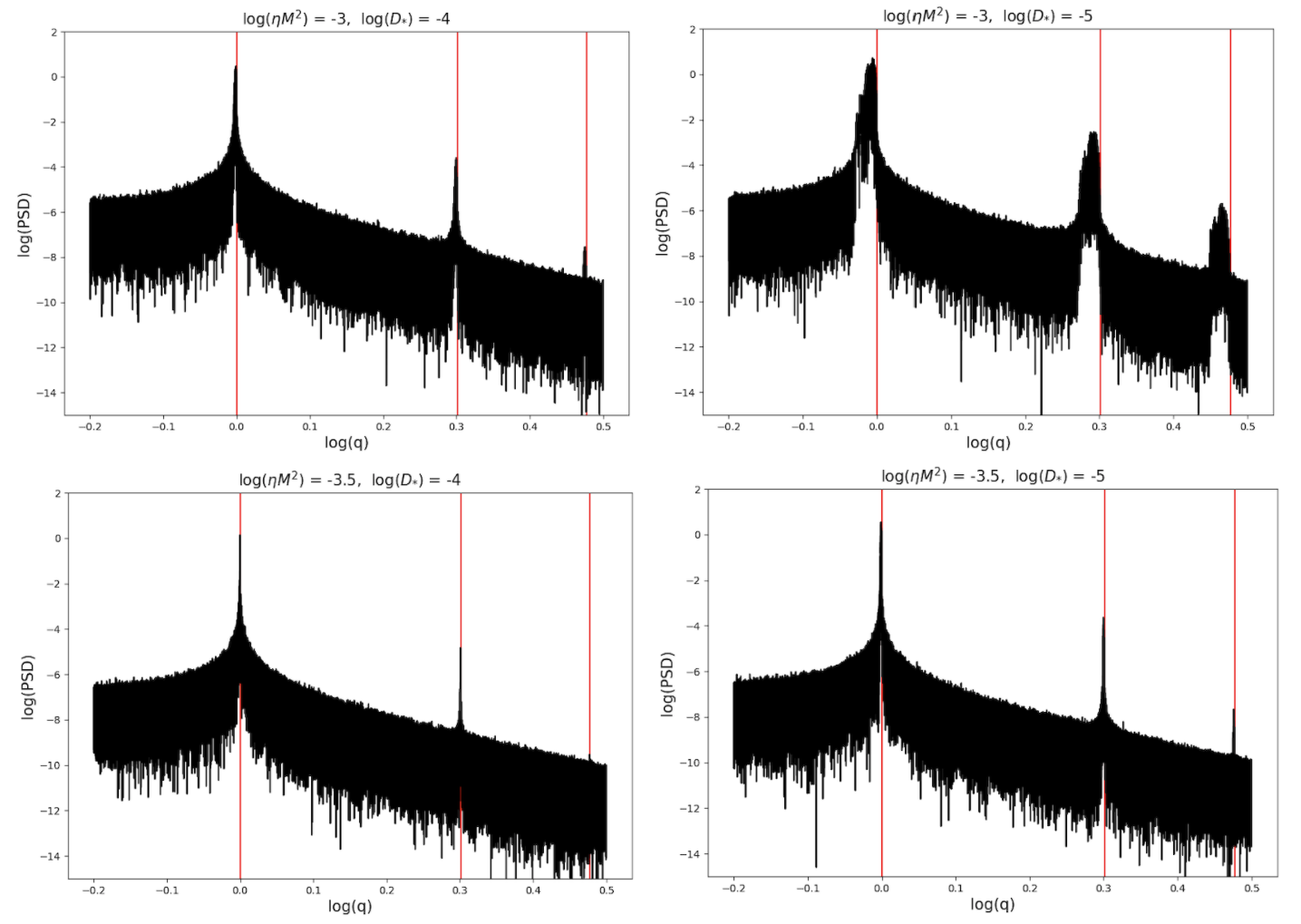}
\caption{ Power spectral density of the amplitude $A$ of the mode, with $q \equiv \omega/\omega_0$ . The two smallest values of the damping parameter that did not produce a blowout (for the fiducial value of $\eta\mathcal{M}^2$) are chosen. The harmonics ($q = 1, 2, 3$) of the unperturbed linear oscillator are indicated by the red lines.}  
\end{figure}

\begin{figure}[tbp]
\plotone{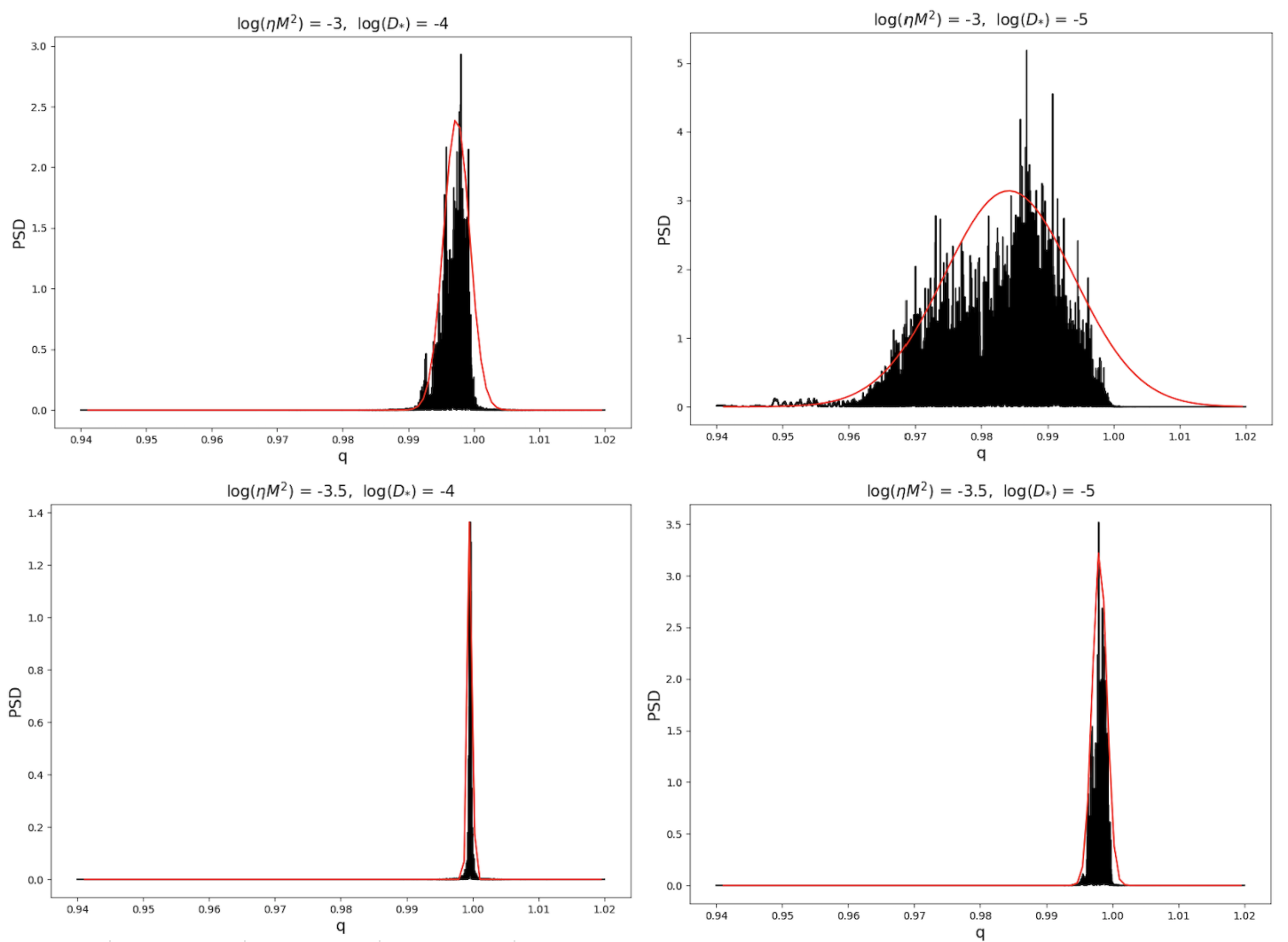}
\caption{Detail of the fundamental mode shown in the previous figure, with a Gaussian fit. Then $\int P_A(q) dq \approx 1.1 P_0 \Delta q$, where $P_0$ is the maximum value of the PSD and $\Delta q$ is the full width at half maximum.}  
\end{figure}

However, it is also seen from Figures 8 and 9 (and their extension for other values of the parameters) that the maximum value of each of the PSDs of the fundamental mode is $P_0 \sim 3$ ,where $\int P_A(q) dq \approx 1.1 P_0 \Delta q $. Therefore, employing equations~(\ref{eq:Pdq}) and (\ref{eq:fP}), we see that visibility of a HFQPO [$P_x(\mbox{max., QPO}) >  P_{x}(\mbox{cont.})$]  requires that 
\begin{equation}
\chi Q\langle A^2 \rangle \gtrsim f_0 P_{x}(\mbox{cont.}) \sim (2 - 300)\times 10^{-4}   \: .        \label{eq:vis}
\end{equation}
Recalling that $\chi \sim (\zeta h/r)^2$, typical values of $\zeta\sim 1$ and $h/r \sim 10^{-2}$ and the requirement $A^2 < 1$ could produce a visible HFQPO if $Q \gg 1$ or $\chi \gg \chi(\mbox{typical}) \sim 10^{-4} $. For instance, \citet{ORM} found that the trapped g-mode leaked into outgoing p-waves in their viscous hydrodynamic simulation. This would increase the value of $\zeta = \Delta r /h$. 

\citet{nw} estimated the contribution of the turbulence to the continuum PSD. Referring to their Figure 1, it is seen that near the frequencies of the HFQPOs in BHBs,
\begin{equation}
f_0 P_{x}(\mbox{turb.}) \approx 0.1 (L/L_{Edd})\mathcal{M}^4 \; . \label{eq:fPturb}
\end{equation}
Then with $ L \lesssim L_{Edd} $ and $ \mathcal{M} \lesssim 0.1 $, we obtain $ f_0 P_{x}(\mbox{turb.}) \lesssim 10^{-5} $ . 
Comparing with the observed values of $f_0 P_{x}(\mbox{cont.})$ above, we see that it is unlikely that turbulence is the major contributor to the observed continuum PSD. This result is consistent with the fact that at lower frequencies, it is also seen from their Figure 1 that the turbulence in the accretion disk at the correspondingly larger radii provides very little of the observed power in the X-ray band. 

\section{Discussion}

A critical feature of the physical conditions that we have investigated is the fundamental difference between the nature of our turbulent damping function $dH/d\tau$ and the `turbulent viscosity' $\nu_t \approx \ell v_{\ell} /3$ (analogous to molecular viscosity) employed in the analysis of the effects of turbulence acting on the rate of shear of a quasi-steady flow \citep{tb} . See in particular \citet[section 11.5]{kfm}. Since the contribution of turbulence to $F_{1}(\tau)$ is a total time derivative, there is equal probability of short-term positive and negative damping. There should be no long-term temporal correlations between the mode and the turbulent eddies.

The physical origin of the damping of the p-modes in the Sun is uncertain \citep{sb}, but it should involve the the coupling to higher frequency p-waves which are damped when their wavelength becomes less than the scale height at the photosphere. This may also be true for our accretion disks. The generation of Alfv\'{e}n waves could also contribute to the damping. \citet{nwbl} found that changes in entropy produced by radiative transfer effects led to growth of modes within the types of accretion disks considered here. 

We hope to consider magnetic forces in the future. In relevant numerical MHD simulations, although the ratio of magnetic to gas pressure is typically a few percent, the ratio of magnetic pressure to the Reynolds stress can be greater than unity \citep{dlof}. In addition, the magnetic forces  can reduce the trapping of the g-modes \citep{fl}, but the amount depends on the relative magnitude of the poloidal and toroidal components \citep{osawl}. 
\citet{rm09} found that the g-mode did not appear in their GRMHD simulations, although the number of orbits may have been insufficient to see growth of the mode. However, \citet{dlof} found that a small eccentricity in accretion disk orbits can excite r(g) - modes to large amplitudes in their MHD simulations. Warps can also excite such modes \citep{k04,k08}. 
 
In the future, we also hope to refine the modeling of the turbulence. One issue is the collective correlated effect of the eddies within the mode. Another is the effects of anisotropy, in particular the stretching of eddies in the $\vec{e} _\phi$ direction. In addition, could (a) the infrequent significant increases of the mode energy or (b) a temporary reduction in the damping or (c) intermittency in the turbulent cascade be relevant to the low duty cycle exhibited by the HFQPOs? 

The approach that we have taken to this problem may be applicable to other physical systems in which an oscillator is coupled to a turbulent environment. 

\begin{acknowledgments}
We thank Nicole Lloyd-Ronning, Jeff Scargle, and James Stone for helpful comments. 
CT acknowledges financial support from the Stanford Physics Department Summer Undergraduate Research Program and NASA award 80NSSC20K0591 to Krista Lynne Smith. 
We would like to thank Stanford University and the Stanford Research Computing Center for providing computational resources on the Sherlock cluster and support. Support was also provided by the Center for Space Science and Astrophysics at Stanford. We thank the referee for helpful comments. 
\end{acknowledgments}

\end{document}